\documentclass{article}
\usepackage{graphicx}
\def\be{\begin{equation}}
\def\ee{\end{equation}}
\def\bea{\begin{eqnarray}}
\def\eea{\end{eqnarray}}
\def\l{\label}
\def\r{{\bf r}}

\def\hahat{\hat{H}}

\def\hahat0{\hat{H}_0}

\def\cos{\hbox{cos}}
\def\sin{\hbox{sin}}

\def\exp{\hbox{exp}}

\def\d{\hbox{d}}
\def\eps{\varepsilon}
\def\epsi{{\cal E}}
\def\siml{\hbox{\kern.1em \lower.6ex \hbox{$\sim$} \kern-1.12em
 \raise.6ex \hbox{$<$} \kern.1em}}
\def\simg{\hbox{\kern.1em \lower.6ex \hbox{$\sim$} \kern-1.12em
 \raise.6ex \hbox{$>$} \kern.1em}}

\begin{document}

\markboth{A.G. Magner, A.I. Sanzhur, and A.M. Gzhebinsky}{ASYMMETRY 
AND SPIN-ORBIT EFFECTS IN BINDING ENERGY
IN THE EFFECTIVE NUCLEAR SURFACE APPROXIMATION
}

\title{ASYMMETRY AND SPIN-ORBIT EFFECTS IN BINDING ENERGY
IN THE EFFECTIVE NUCLEAR SURFACE APPROXIMATION
}

\author{A.G. Magner\footnote{
magner@kinr.kiev.ua}, A.I. Sanzhur, A.M. Gzhebinsky\\
{\em Institute for Nuclear Research, Kyiv 03680, Ukraine}}

\maketitle

\begin{abstract}
Isoscalar and isovector particle
densities are 
derived analytically by using 
the approximation of a sharp edged nucleus within the 
local energy density approach with the proton-neutron 
asymmetry  and  
spin-orbit effects. 
Equations for the effective nuclear-surface shapes as collective variables are
derived up to the higher order corrections 
in the form of the macroscopic boundary conditions. The analytical
expressions for the isoscalar and isovector tension coefficients of
the nuclear surface binding energy and 
the finite-size
corrections to the $\beta $ stability line are obtained.
\end{abstract}

\section{Introduction}

The simple and accurate solution of some problems involving the particle
density distributions uses the nuclear effective surface (ES) 
approximation\cite{strtyap,tyapin,strmagbr,strmagden}. 
It exploits 
the property of saturation 
of the nuclear matter 
and a narrow diffuse-edge region 
in finite nuclei. 
The ES is  defined 
as the location of points of the  
density gradient maximum. 
The coordinate system related locally to the ES  
is specified by
a distance $\xi$ from the 
given point to the surface and tangent coordinate $\eta$ (see Fig.~1). 
The variational condition of the nuclear energy minimum
 at fixed other integrals of motion within
the local energy density theory, in particular, the extended Thomas-Fermi 
(ETF) approach\cite{brguehak,brbhad} is simplified much in the 
$\xi,\eta$ coordinates for 
any deformations by using expansion 
in small parameter $a/R \sim A^{-1/3} \ll 1$ 
for heavy enough nuclei ($a$ is the diffuse edge thickness of the nucleus, 
and $R$ its mean curvature radius). 
The accuracy of the ES  approximation in the ETF approach
was checked\cite{strmagden} 
by comparing results 
of the  Hartree-Fock (HF)
and ETF theories
based on 
Skyrme forces\cite{brguehak,chaban}  
without spin-orbit and asymmetry terms.
Within the ES approximation a rather reasonable agreement 
of the calculations with the 
experimental data on a mean particle-number dependence
on the excitation energies and reduced transition 
probabilities of the low-lying collective states of non-magic nuclei
was found\cite{lowcolldyn}.
In the present work, we extend the ES approach\cite{strmagden} 
taking into account the spin-orbit and the asymmetry effects in nuclei.

\begin{minipage}{0.45\textwidth}
\includegraphics[width=\textwidth,clip]{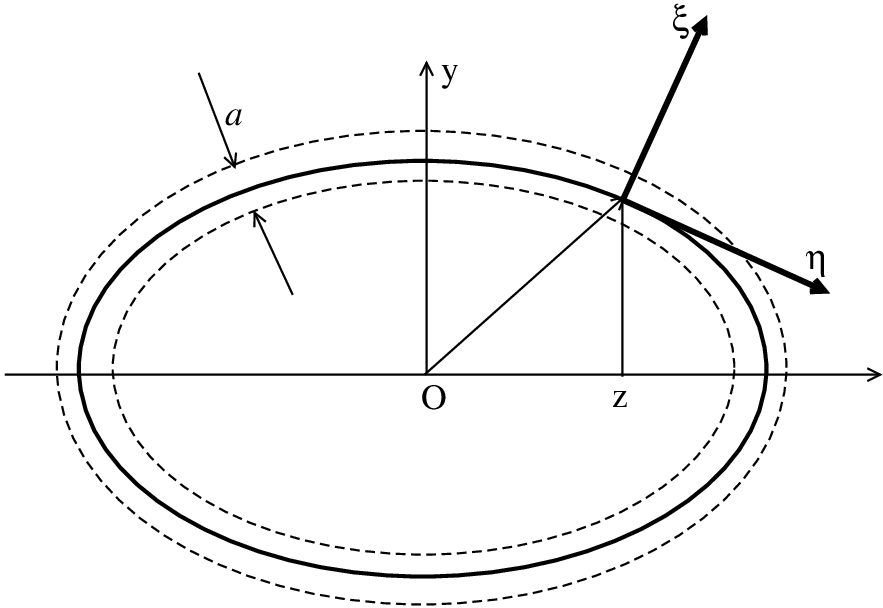}
\end{minipage}
~
\begin{minipage}[t]{0.45\textwidth}
Fig.~1: ES and local  $\xi,\eta$ coordinates. 
The profile function $y=Y(\eta)$ in cylindrical $y,z$ 
coordinates
is shown schematically by
thick solid curve \cite{strtyap,strmagbr};  $a$ is the 
parameter of the nuclear diffuse edge. 
\end{minipage}

\section{LOCAL ENERGY DENSITY AND 
CONSTRAINTS}
\l{enden}

We begin with the nuclear energy $E$ within a local energy 
functional approach\cite{brguehak,chaban}:
\bea\l{energytot}
E&=&\int \d \r\; \epsi [\rho_{+}(\r),\rho_{-}(\r)],\qquad
\epsi\left(\rho_{+},\rho_{-}\right) \approx 
- b_v \rho_{+} + \frac{b_{sym}}{2} {\cal X}^2 \rho_{+}
 + \frac{e}{4}\left(1-{\cal X}\right) \overline{\Phi} \rho_{+}
\nonumber\\
&+&
\rho_{+} \left[\eps_{+}(\rho_{+}) - \eps_{-}\right]
+ \left({\cal A} +{\cal B} \rho_{+} +\frac{\Gamma}{4 \rho_{+}}\right) 
\left(\nabla \rho_{+}\right)^2  
+ {\cal A}_{-} \left(\nabla \rho_{-}\right)^2,
\eea
where $\epsi [\rho_{+}(\r),\rho_{-}(\r)]$ 
is the energy density as a function
of the isoscalar $\rho_{+}$ and isovector $\rho_{-}$ particle densities.
It overlaps approximately most of the
realistic Skyrme forces\cite{chaban}.
$\rho_{\pm}=\rho_n \pm \rho_p$,
$~{\cal X}=(N-Z)/A$,
$~N=\int \d \r \rho_n$, $~Z=\int \d \r \rho_p$,
$~A=N+Z$,  $~\Phi$ is the Coulomb potential and $\overline{\Phi}$ is
its average up to a small exchange component 
 \cite{strtyap,tyapin,bormot}. As usually, ${\cal E}$ of (1) contains
the volume, and the surface terms without and with the gradient density terms
 \cite{strtyap,strmagbr,strmagden}, 
$b_v$ =16 MeV is the separation energy per particle 
and  $~b_{sym}$=60 MeV is the symmetry energy constant of the nuclear matter.
The semiclassical $\hbar$ corrections appear through 
$~\Gamma=\hbar^2/18 m$ in the ETF kinetic energy 
density\cite{brguehak,brbhad}, $m$ is the nucleon mass. 
In (\ref{energytot}), we have neglected relatively small isovector 
(spin-orbit and semiclassical) corrections.
The isoscalar 
surface energy density part, independent of the density gradient terms,
 is determined by the function  
$\varepsilon_{+}(\rho_{+})$ which satisfies
the saturation condition:
\be\l{satcond}
\eps_{+}(\overline{\rho})=0, \qquad\qquad
\d\eps_{+}(\overline{\rho})/\d\rho_{+} =0, 
\ee
where $\overline{\rho}=3/4\pi r_0^3 \approx$ 0.16 fm$^{-3}$ 
is the density of the infinite nuclear matter,
$r_0 = R/ A^{1/3}$ is constant independent of $A$.
For the isovector component one has   
\be\l{epsminus}
\eps_{-}=\frac{b_{sym} }{2}\;\left({\cal X}^2 -
\rho_{-}^2/\rho_{+}^2\right) 
- \frac{e}{4}\; 
 \left[\left(1-\rho_{-}/\rho_{+}\right)\;\Phi -\left(1 - {\cal X}\right)\;
\overline{\Phi}\right].
\ee
The spin-orbit gradient terms in (\ref{energytot}) are defined with a
constant: 
${\cal B} = -9m W_0^2/16 \hbar^2$,
$W_0$=100 - 130~ MeV fm$^{5}$ (see refs. 5,7).

From the condition of the energy $E$ (\ref{energytot}) minimum 
together with the constraints
for the fixed particle number $A$, neutron excess $N-Z$, and deformation $Q$
of the nucleus\cite{strtyap,tyapin,strlyaschpop}:
\be\l{constraintsATQ} 
A = \int \d \r\; \rho_{+}(\r),\qquad\,\,\, 
N-Z = \int \d \r\; \rho_{-}(\r),\qquad\,\,\,
Q = \int \d \r\; \rho_{+}(\r)\; q(\r),
\ee
one arrives at the variational Lagrange equations:
\be\l{lagrangeqt}
\frac{\delta \epsi}{\delta \rho_{+}} -\lambda_{+} -\lambda_Q\; q 
=0,\qquad\qquad
\frac{\delta \epsi}{\delta \rho_{-}} - \lambda_{-} =0.
\ee 
Here, $\lambda_{+}$, $\lambda_{-}$ and $\lambda_{Q}$ are 
the corresponding Lagrange
multipliers where $\lambda_{+}$ and $\lambda_{-}$ are the 
isoscalar and isovector 
chemical potentials, respectively.

\section{ISOSCALAR AND ISOVECTOR PARTICLE DENSITIES}
\l{isivden}

{\it In the nuclear volume}, up to the second order in $\rho_{+} - 
\overline{\rho}$ one gets\cite{strtyap,strmagbr,strmagden}:
\be\l{epsvol}
\eps_{+}(\rho_{+})=\frac{K}{18\; \overline{\rho}^2}\left(\rho_{+} - 
\overline{\rho}\right)^2,
\qquad\qquad \epsilon(w)=\frac{\eps_{+}}{b_v}=(1-w)^2 , \qquad\qquad 
w=\frac{\rho_{+}}{\overline{\rho}}.
\ee
where $K$ is the 
incompressibility of the infinite nuclear matter.
From the Lagrange equations (\ref{lagrangeqt}) one finds
for the volume densities $\rho_{\pm}^{(v)}$ :
\be\l{rhominusvol}
\rho_{+}^{(v)} \approx 
\overline{\rho}\left(1+\frac{9 \Lambda_{tot}^{(+)}}{K}\right),
\qquad\qquad \rho_{-}^{(v)} \approx 
\overline{\rho} \left({\cal X}+
\frac{\Lambda_{tot}^{(-)}}{b_{sym}} \right).
\ee
Small finite-size corrections of the order of $a/R \sim aH$ 
($H$ is a mean ES curvature)
are determined by the surface components
of the corresponding chemical potentials:
\bea\l{lambdat}
\Lambda_{tot}^{(+)} &=&
\lambda_{+} +  b_v -\frac{b_{sym}}{2}{\cal X}^2 -
 \frac{e}{4}\;\overline{\Phi} + 
\lambda_Q\; q(\r)\sim aH \sim a/R \sim A^{-1/3}, 
\nonumber\\
\Lambda_{tot}^{(-)}&=&\lambda_{-} -b_{sym} {\cal X}+
\frac{e}{4}\;\overline{\Phi}\sim aH.
\eea

{\it For the dimensionless isoscalar density} 
$w(x)$
(\ref{epsvol}), from the first equation (\ref{lagrangeqt}), {\it up 
to the leading order} in $a/R$
one obtains the ordinary first-order differential equation:
\be\l{yeq0plus} 
\frac{\d w}{\d x} =-w\;\sqrt{\frac{\epsilon(w)}{w + \beta w^2 + \gamma}},
\qquad\qquad
x=\frac{\xi}{a},\qquad 
a=\sqrt{\frac{{\cal A}\; \overline{\rho}\; K}{18\; b_v^2}},
\ee
where 
$\beta={\cal B}\overline{\rho}/{\cal A}$, 
$\gamma=\Gamma/4 \overline{\rho}{\cal A}$.
By differentiating equation (\ref{yeq0plus}) 
one finds the boundary condition from the definition of the ES: 
$\partial^2 w/\partial x^2 =0$ at $x=0$ ($\xi=0$),
\be\l{boundcond}
\left(w_0 -\beta w_0^2 - \gamma \right) \epsilon(w_0) 
+ w_0\left(w_0 + \beta w_0^2 + \gamma\right) 
\left(\frac{\d \epsilon(w)}{\d w}\right)_{w=w_0}=0,  
\ee 
together with the condition
 of the exponentially vanishing the density outside the nucleus: 
$w \sim \exp(-x)=\exp(-\xi/a)$.
Solving the problem (\ref{yeq0plus}), (\ref{boundcond}), one arrives at the
solution in the inverse form $x(w)$:
\be\l{ysolplus}
x=-\int_{w_0}^{w}\frac{\d \tau}{\tau} \sqrt{\frac{\tau +\beta \tau^2 
+ \gamma}{\epsilon(\tau)}}.
\ee
With the quadratic approximation (\ref{epsvol}) for $\epsilon(w)$
one gets the analytical solutions in terms of the 
algebraic, trigonometric and
logarithmic functions.
For $\beta=\gamma=0$ it simplifies to 
$~w(x)=\tanh^2\left[(x-x_0)/2\right]$ for
$x\leq x_0=2{\rm arctanh}\left(1/\sqrt{3}\right)$ 
and zero for $x$ outside the nucleus\cite{strmagden}.
In Fig.~2 (left) the influence of the
 semiclassical correction to $w(x)$ is shown by comparing
the $\gamma=0$ (dashed line) and the ``exact'' (thin solid line) 
cases. This correction is small everywhere, 
besides the quantum tail outside 
the nucleus for $x \simg 1$. Almost the same results one obtains for
the SkM$^*$ and SLy7 forces\cite{chaban}. One should also notice 
a rather big effect of the 
spin-orbit interaction as compared to the simplest analytical solution at
$\beta=\gamma=0$. 
We found also a good convergence of the expansion of the
$\epsilon(w)$ in powers of $1-w$ in the density solution
(\ref{ysolplus}) by comparing the exact
numerical function\cite{chaban} $\epsilon(w)$ 
 to its approximate solution (\ref{epsvol}).
The agreement is within a precision of the line thickness. 
Fig.~2 (right) presents a weak sensitivity of the isoscalar solution
$w(x)$ (\ref{ysolplus}) on 
the different Skyrme forces with (\ref{epsvol}) for $\epsilon(w)$. 

~\vspace*{0.78cm}

\includegraphics[width=0.46\textwidth]{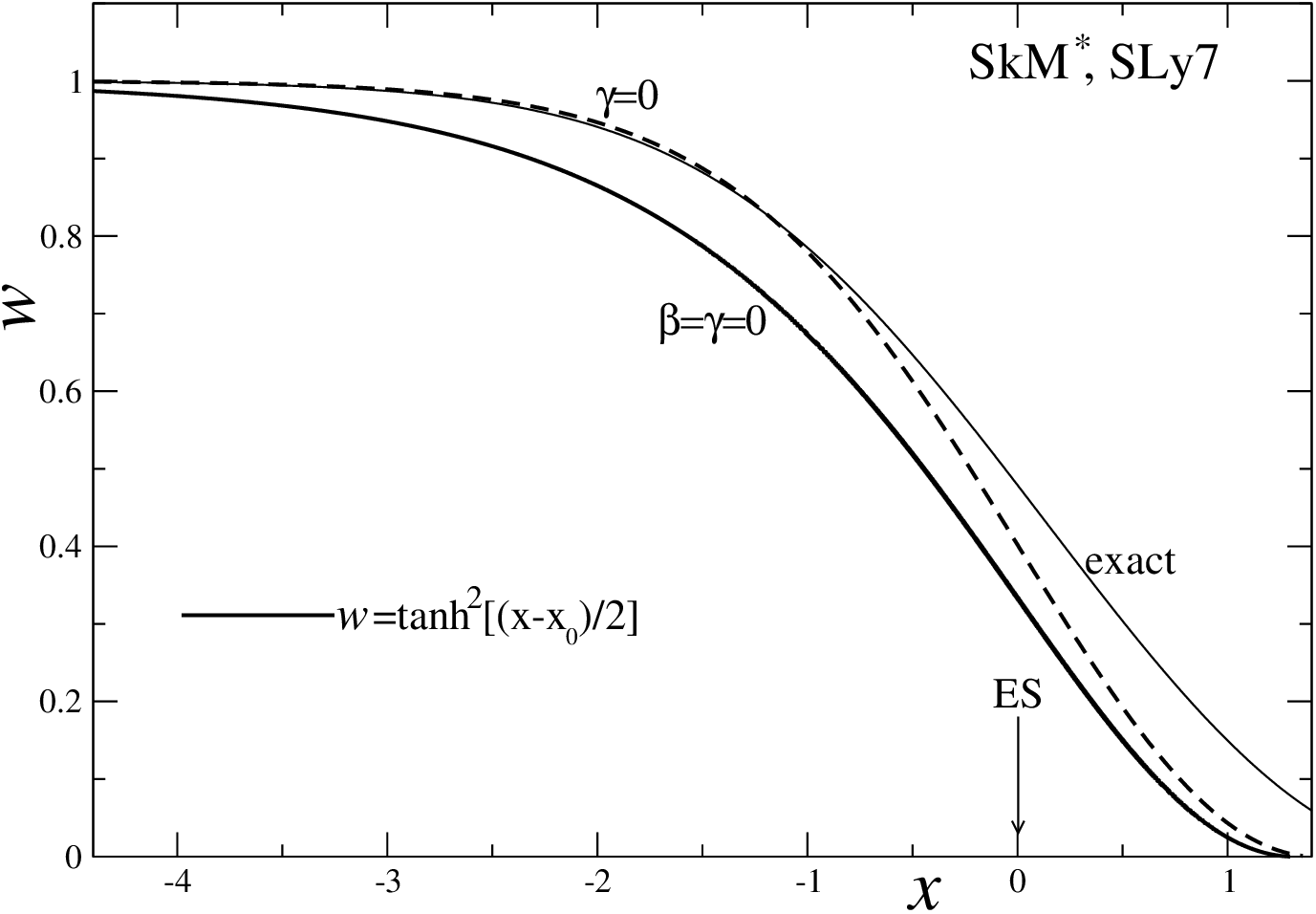}
\hspace{0.2cm}
\includegraphics[width=0.46\textwidth]{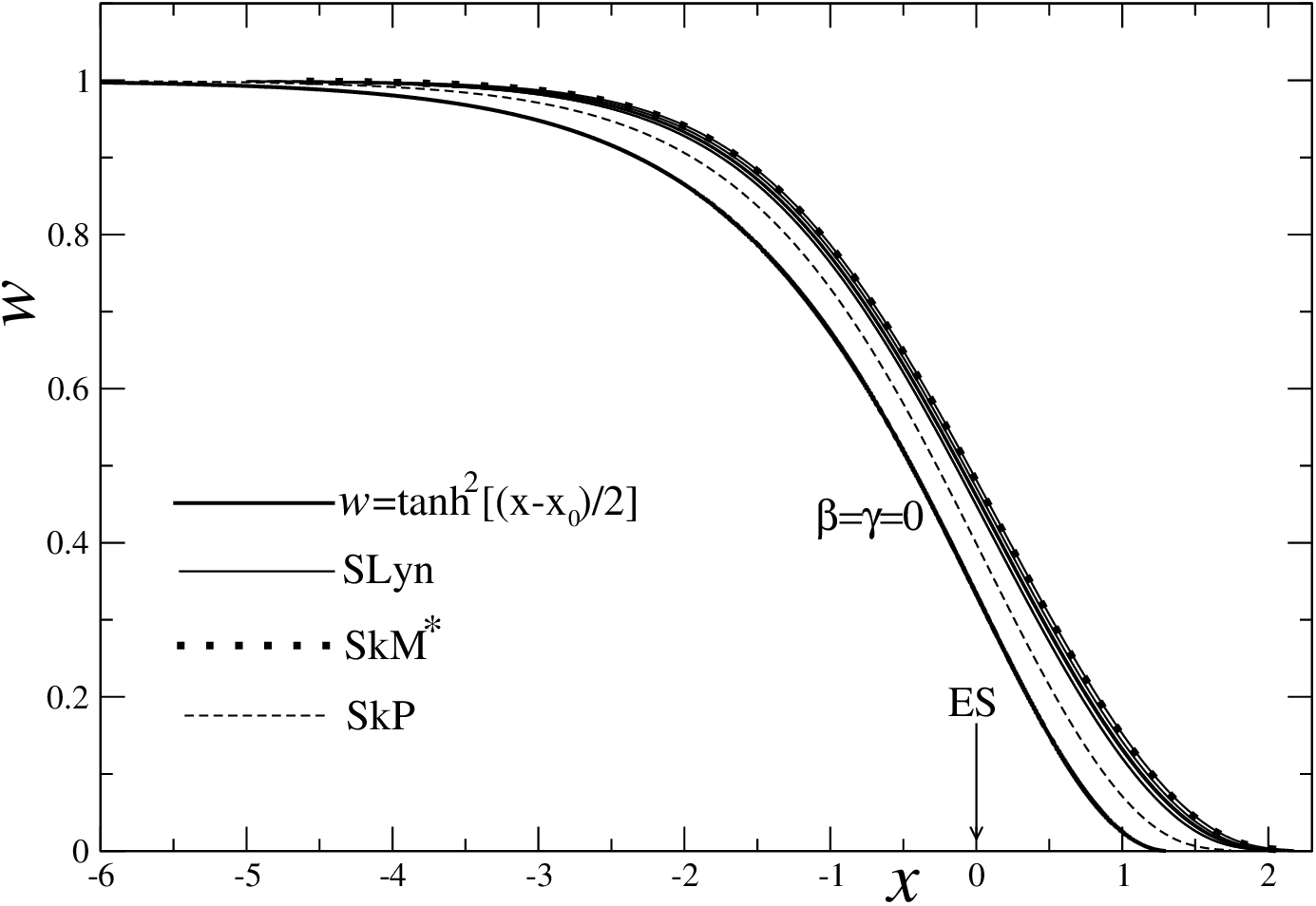} 

\vspace{0.32cm}
Fig.~2: Density $w(x)$ (\ref{ysolplus}) (left) as a 
function of $x=\xi/a$ and its
comparison (right) for several Skyrme forces\cite{chaban} 
($n=$5-7,230a and 230b in SLyn) for $\epsilon(w)$ (\ref{epsvol}).
\vspace{0.2cm}

{\it For the isovector density up to the leading order} in $a/R$, 
after simple transformations one finds the equation and the
 boundary condition in the form
\be\l{yeq0minus} 
\frac{\d w_{-}}{\d w} =c_{sym}
\sqrt{\frac{1+ \beta w}{\epsilon(w)}}
\;\sqrt{1 -\frac{w_{-}^2}{w^2}}, \qquad w_{-}(1) =1,\qquad
 w_{-}=\frac{\rho_{-}}{\rho_{-}^{(v)}} \approx 
\frac{\rho_{-}}{\overline{\rho}{\cal X}},
\ee
where  $c_{sym}=a \;\sqrt{-b_{sym}/2 \overline{\rho}\;{\cal A}_{-}}$.
Up to the leading order of the ES approximation in 
$a/R$, one obtains the analytical solution through the expansion in powers of
$1-w$,
\be\l{ysolminus}
w_{-} = 
 w\;\cos \left[u(w)\right],\qquad
u(w)=\frac{1-w}{c_{sym}\sqrt{1+\beta}} \left[1
+ \frac{1-w}{c_{sym}\left(1+\beta\right) + \beta/2}\right].
\ee
The dependence of the dimensionless
isovector density $w_{-}(x)$ (\ref{ysolminus})
on the semiclassical and spin-orbit effects 
versus
the corresponding results for the density $w(x)$
(\ref{ysolplus}) are shown in Fig.~3 (left). A weak sensitivity
of the dimensionless isovector density $w_{-}$ 
on the choice of the Skyrme 
forces is seen in Fig.~3 (right). 

\section{  
ES EQUATIONS AND LDM 
BOUNDARY CONDITIONS}
\l{esldm}
{\it For more exact isoscalar particle density} we calculate the
main terms of higher order in the parameter $a/R$ in the first equation 
(\ref{lagrangeqt}). Integrating this equation over the ES
in normal-to-surface $\xi$ direction and using the 
equation (\ref{yeq0plus})
up to the leading order in $a/R$, one arrives at the differential equation 
\bea\l{boundcondPplus}
P_{+}\Big|_{ES} &=&
P_s^{(+)},\qquad
P_{+}=\overline{\rho}\; \Lambda_{tot}^{(+)}=
\left\{\frac{K}{2 \overline{\rho}}\;\left(\rho_{+} -
\overline{\rho}\right)
\left[1 + \frac{3}{2 \overline{\rho}}\;\left(\rho_{+} -
\overline{\rho}\right)\right] 
+ \frac{b_{sym}}{2 \overline{\rho}^2}\;
\left[\left(\rho_{-}\right)^2\right.\right.
\nonumber\\
&-& \left.\left.\overline{\rho}^2 \;{\cal X}^2\right]
+ \frac{e \overline{\rho}}{4}\;
\left[\frac{\d}{\d \rho_{+}}\;\left((\rho_{+}-\rho_{-})\Phi\right)
- \overline{\Phi}\right]\right\}^{(v)},\qquad
P_s^{(+)}=2 \sigma_{+} H.
\eea
This equation can be considered with respect to the unknown sought 
 profile shape $y=Y(\eta)$ of the ES in
the cylindrical 
coordinates $y,z$ with the symmetry axis $z$ (see Fig.~1) 
through the  
curvature $H=(1/R_1+1/R_2)/2$ in terms of the main ES curvature 
radii\cite{strtyap,strmagbr} 
\be\l{curvature}
R_1={\cal L} Y(\eta), \qquad
R_2=-{\cal L}^3/\left(\partial^2 Y/\partial \eta^2\right),
\qquad
{\cal L}= \left[1+ \left(\partial Y(\eta)/\partial \eta\right)^2
 \right]^{1/2}.
\ee

\vspace{0.4cm}
\includegraphics[width=0.45\textwidth]{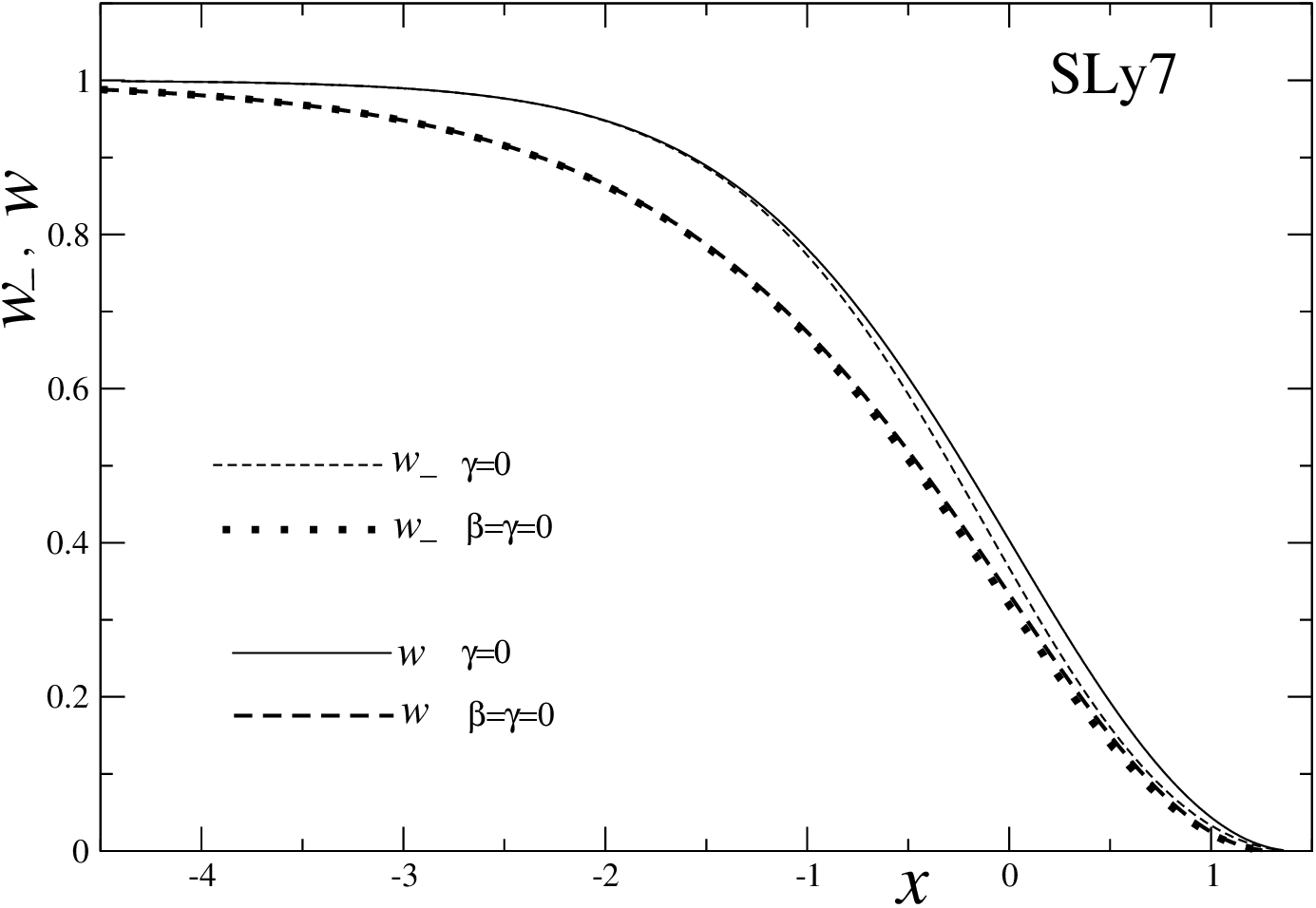}
\hspace{0.5cm}
\includegraphics[width=0.45\textwidth]{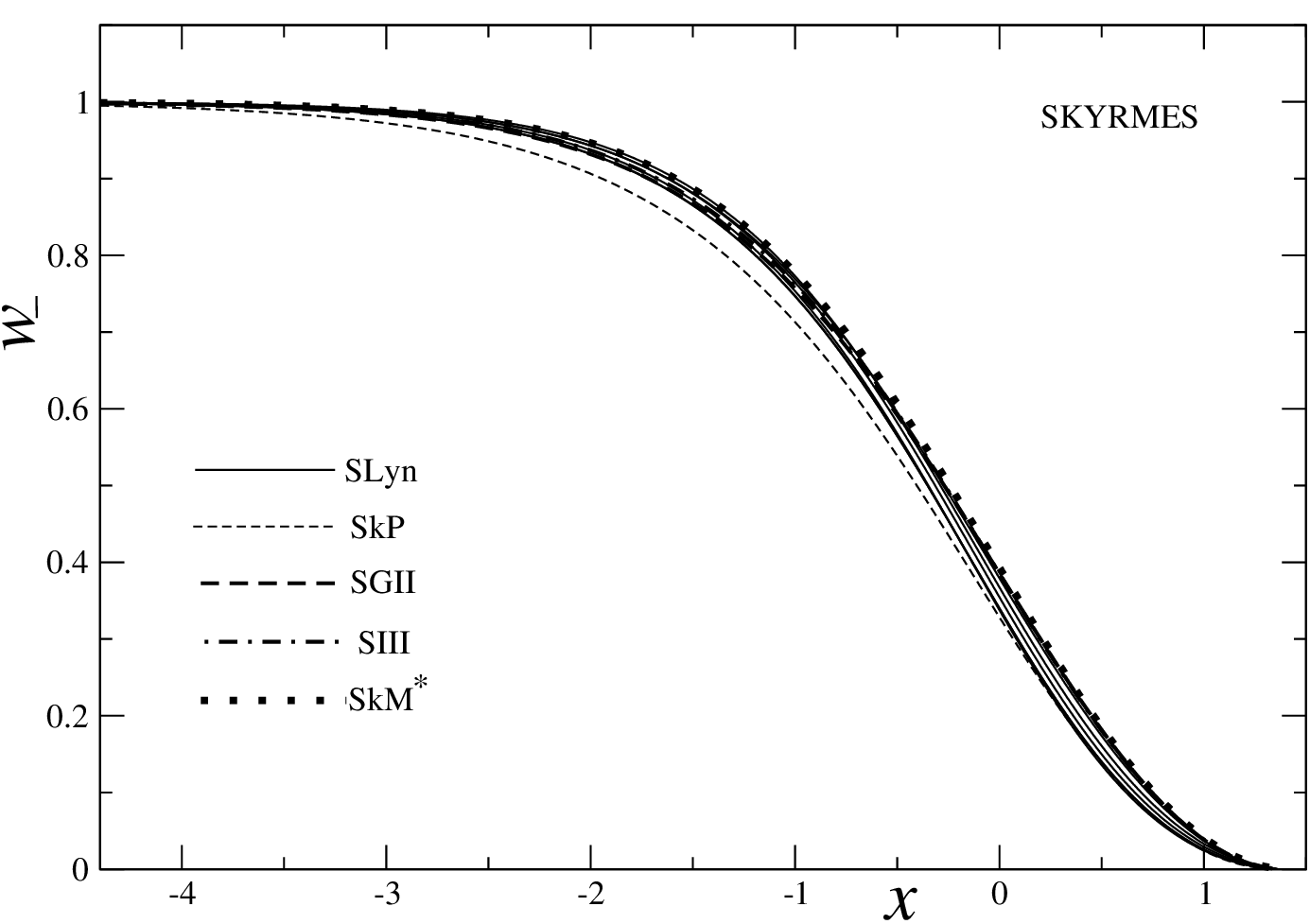}

\vspace{0.5cm}
Fig.~3: Isovector density 
$w_{-}(x)$ (\ref{ysolminus}) compared to the isoscalar one 
$w(x)$  (\ref{ysolplus})
as a function of $x=\xi/a$ within the  
approximation (\ref{epsvol}) to 
$\epsilon(w)$ calculated
for the same SLy7 forces (left) and $w_{-}(x)$
for  several Skyrme forces\cite{chaban} (right). 

\vspace{0.4cm}
\noindent
Eq.~(\ref{boundcondPplus}) 
is associated with the macroscopic boundary 
condition\cite{bormot,magstr,kolmagsh} 
with the isoscalar 
capilliary surface pressure $P_s^{(+)}$ which is 
proportional to the surface 
tension coefficient  $\sigma_{+}$:
\be\l{sigmaplus}
\sigma_{+} 
\approx
2\int_{-\infty}^{\infty} \d \xi \left({\cal A} +{\cal B} \rho_{+} +
\frac{\Gamma}{4 \rho_{+}}\right)\; 
\left(\frac{\partial \rho_{+}}{\partial \xi}\right)^2. 
\ee 

{\it For more exact isovector particle density} similarly one   
obtains the isovector macroscopic boundary condition\cite{kolmagsh},
\be\l{boundcondPminus}
P_{-}\Big|_{ES}= P_s^{(-)},\qquad
P_{-}=\overline{\rho} {\cal X} \Lambda_{tot}^{(-)}=
{\cal X}\;\left[b_{sym}\left(\rho_{-}/\overline{\rho}
-{\cal X}\right) -
\frac{e}{4} \left(\Phi - \overline{\Phi}\right)\right]^{(v)},
\ee
with the isovector surface pressure 
\be\l{sigmaminus}
P_s^{(-)}=2 \sigma_{-} H, \qquad\sigma_{-} \approx
2 {\cal A}_{-} \int_{-\infty}^{\infty}\d \xi \; 
\left(\frac{\partial 
\rho_{-}}{\partial \xi}\right)^2, 
\ee
where $\sigma_{-}$ is the isovector tension coefficient.

\section{SURFACE ENERGY}
\l{surfent}

The nuclear energy $E=E_v+E_s$ (\ref{energytot}) in the ES
approximation is split into the
volume, $E_v=-b_v\; A +b_{sym} (N-Z)^2/2A + 
eZ \overline{\Phi}/4$, and the surface terms:
\be\l{EvEs}
 E_s = 
\sigma\;S=
\left(b_s^{(+)} + b_s^{(-)}\right) S/4\pi r_0^2,\qquad
\sigma=\sigma_{+} + \sigma_{-},\qquad
b_s^{(\pm)}= 4 \pi r_0^2 \; \sigma_{\pm},
\ee 

\noindent
where $S$ is the surface area of the ES. 
The energy $E_s$ (\ref{EvEs}) is determined by the sum of the
isoscalar $b_{s}^{(+)}$ and isovector $b_{s}^{(-)}$
surface energy constants. These constants are 
proportional to the same tension 
coefficients $\sigma_{\pm}$ which appear in (\ref{sigmaplus}) and
(\ref{sigmaminus}) and expressed 
through the surface pressures in 
(\ref{boundcondPplus}) and (\ref{sigmaminus}), respectively,
\be\l{bsplus} 
b_s^{(+)}=
 \frac{54 a b_v^2}{K r_0} 
\;\int_0^1 \d w \;\sqrt{\left(w +\beta w^2 + \gamma\right) \epsilon(w)},
\ee
\be\l{bsminus} 
b_s^{(-)}=
108 \alpha_{-} {\cal X}^2 \frac{a b_v^2}{K r_0}\; 
\int_{0}^{1} \d w\; \frac{\sqrt{w} (1-w)}{\sqrt{1+\beta w}}\;
\left\{\cos [u(w)]
- w\; \sin [u(w)]\; u'(w)\right\}^2,
\ee
where $\alpha_{-}={\cal A}_{-}/{\cal A}$ and  
see (\ref{ysolminus}) for $u(w)$.
Simple expressions for constants (\ref{bsplus}) and 
(\ref{bsminus})  in terms of
the algebraic and trigonometric functions can be easily
obtained by calculating 
explicitly the integrals over $w$ in the above equations
 with $\epsilon(w)$ taken from eq.(6).
Neglecting relatively small spin-orbit terms and semiclassical corrections
one finds 
the approximate relationship between the isovector and the isoscalar
energy constants,
$b_s^{-} \approx \alpha_{-}\;{\cal X}^2 b_s^{+}$.

In Table~1 the analytical, $b_{s, an}^{(+)}$
with the approximated (\ref{epsvol}) for $\epsilon(w)$, the
numerical, $b_{s, num}^{(+)}$ with the exact $\epsilon(w)$,
and $b_{s}^{(+)}$ from ref.7 are shown for all Skyrme forces.
One can see a very good agreement between all these calculations, 
besides of SIII.
Modula of the isovector constants for the Lyon Skyrme forces SLyn\cite{chaban} 
are much larger than for other ones. 
The precision of the spin-orbit and semiclassical 
terms of the isovector energy density part  
(\ref{energytot})
is not enough accurate for all considered 
Skyrme interactions as
 the isovector 
surface tension $\sigma_{-}$ which appears in the
surface energy (\ref{EvEs}) becomes inconsistent with that of
the capilliary isovector pressure (\ref{boundcondPminus}), 
 (\ref{sigmaminus}).  
These terms can be improved by fitting to
more detailed experimental information.

{\it The $\beta$-stability line} is determined by 
the equivalence of the neutron 
and proton  chemical potentials,
$\lambda_{-}= \lambda_n -\lambda_p % b_{sym} \;{\cal X} - 
%\frac{e}{4} \;\overline{\Phi}_0\; \left(1-{\cal X}\right) 
%+\Lambda_{tot}^{(-)} 
=0$, and $\Lambda_{tot}^{(-)}=
2b_{s}^{(-)} H/4 \pi r_0^2\overline{\rho}{\cal X} %\approx 
%2 b_s^{(-)}/3 {\cal X} A^{1/3}
$, according to (\ref{lambdat}), (\ref{boundcondPminus}) and
(\ref{sigmaminus}).
With {\it the finite-size
correction},
one obtains  
${\cal X}
\approx
{\cal X}_0 \left(1-2 b_s^{(-)} r_0 H/3 b_{sym} {\cal X}_0^2\right)$,
where ${\cal X}_{0} \approx 3 A^{2/3} e^2/10 r_0 b_{sym}$
is the leading term\cite{bormot}. 
\hspace{1cm}
\begin{table}[pt]
{\begin{tabular}{|c|c|c|c|c|c|c|c|c|}
\hline
 & SkM$^*$ & SIII & SGII & SLy230a & SLy230b & SLy4 & SLy6 & SLy7\\
\hline
$b_{s,an}^{+}$  & 17.1&  11.8 & 15.3 
& 17.3 & 17.5 & 17.5 & 17.5 & 15.9  \\
$b_{s,num}^{+}$  & 18.5&  11.6 & 16.5
& 18.5 & 18.7 & 18.7 & 18.7 & 17.0  \\
$b_{s}^{+}$  & 16.0&  17.0 & 14.8 
& 16.9 & 16.7 & 18.1 & 17.4 & 17.0  \\
$b_{sym}$  & 60.1&  56.3 & 53.7 
& 63.9 & 64.0 & 64.0 & 63.9 & 63.9  \\
$b_{s}^{-}/{\cal X}^2$  & -3.23& -3.72& -1.08 
& -7.61 & -26.3  & -26.3 & -15.7 & -10.5  \\
\hline 
\end{tabular}
}

\vspace{0.5cm}
Table~1: The isoscalar (\ref{bsplus}) and 
isovector (\ref{bsminus}) energy surface constants  $b_s^{\pm}$
with  Skyrme parameters \cite{chaban}. 
\end{table}

\noindent

\section{CONCLUSIONS}
\l{concl}

The asymmetry and spin-orbit 
terms of the energy density 
within the ETF with Skyrme forces
were taken into account  analytically by using expansion in 
$a/R \ll 1$ of the ES at any deformation. 
We derived the ordinary first-order equations for  
the isoscalar 
and isovector particle densities giving 
simple analytical solutions.   
When  higher order terms are taken into account one gets
equations for the moving 
ES in terms of the macroscopic
 boundary conditions. 
Expressions for the isoscalar and isovector 
tension coefficients $\sigma_{\pm}$ in the surface energy 
were found as those of the 
macroscopic capilliary pressures of the Fermi liquid edge in these 
boundary conditions. 
A simple approximate relation of $b_s^{(-)}$ to $b_s^{(+)}$ and
the finite-size ES correction to the
 $\beta$-stability condition were obtained. Our approach might be helpful
as a macroscopic part of the 
nuclear collective dynamical macro-micromodels
\cite{magstr,yaf2,strutmagbrpr}
based on the ETF (or LDM) for a ``macro'' 
component\cite{lowcolldyn}, 
 and the semi-microscopic approach of 
the  Strutinsky shell correction method\cite{brbhad,brdamjen}  
for study of the low-lying collective excitations and 
fission processes.

\section*{Acknowledgements}

Authors thank Profs. V.M. Kolomietz, K. Pomorski,
H.J. Krappe, P. Ring, V.O.~Nesterenko, P. Danielewicz, J. Kvasil, and G. Colo 
for many useful discussions.

\end{document}